\documentclass[doublecol]{epl2} 

\usepackage{amssymb}
\usepackage{amsfonts}
\usepackage{amsmath}
\usepackage{amsthm}
\usepackage{bbm}

\title{Stability of amplitude chimeras in oscillator networks}
\shorttitle{Stability of amplitude chimeras} 

\author{L. Tumash\inst{1} \and A. Zakharova\inst{1} \and J. Lehnert\inst{1} \and W. Just\inst{2} \and E. Sch{\"o}ll\inst{1}}
\shortauthor{L. Tumash \etal}

\institute{                    
  \inst{1}Institut f{\"u}r Theoretische Physik, Technische Universit\"at Berlin, Hardenbergstra\ss{}e 36, 10623 Berlin, Germany\\
  \inst{2}Queen Mary, University of London, School of Mathematical Sciences, Mile End Road, London E1 4NS, UK\\
}

\abstract{We show that amplitude chimeras in ring networks of Stuart-Landau oscillators with symmetry-breaking nonlocal coupling represent saddle-states in the underlying phase space of the network. 
Chimera states are composed of coexisting spatial domains of coherent and of incoherent oscillations. We calculate the Floquet exponents and the corresponding eigenvectors in dependence upon the coupling strength and range, and discuss the implications for the phase space structure. The existence of at least one positive real part of the Floquet exponents indicates an unstable manifold in phase space, which explains the nature of these states as long-living transients. Additionally, we find a Stuart-Landau network of minimum size $N=12$ exhibiting amplitude chimeras.}

\begin{document}

\maketitle

\section{Introduction}
The dynamical state of networks of homogeneously coupled identical elements can show a peculiar behavior by self-organizing into two spatially separated domains with dramatically different behavior, e.g. a spatially coherent and a spatially incoherent region. This phenomenon was named \textit{chimera state} by Abrams and Strogatz \cite{ABR04} after the Greek fire-breathing monster, whose body consists of  different animals. These hybrid states were discovered for phase oscillators in the early 2000's by Kuramoto  and Battogtokh \cite{KUR02a}. They observed a spontaneous breakup of the system into spatially coexisting synchronized and desynchronized domains with respect to the phase. 

Chimera states have possible applications to neural activity \cite{LAI01,HIZ16}, heart fibrillation \cite{DAV92a} and social systems\cite{GON14}. Chimera states were also associated with such phenomena as epileptic seizure \cite{ROT14} and unihemispheric sleep, which has been detected for some sea mammals and birds \cite{RAT00}. These creatures can sleep with only one half of their brain  while the other half remains awake. For instance, this enables sleeping dolphins to detect predators and migrant birds can travel for hundreds of kilometers without having a break \cite{RAT16}. Recently, unihemispheric sleep has been found also for humans \cite{TAM16}.

Chimera states were initially found for coupled phase oscillators, where coherence is related to 
phase- and frequency-locked oscillators and incoherence is associated with
drifting oscillators. Since then numerous chimera patterns have been shown in a variety of systems of different nature including coupled amplitude and phase dynamics \cite{PAN15,KEM16,SCH16b}. Recently, a special type of chimera state has been discovered where coherence and incoherence occur with respect to the amplitude of the oscillators while all the elements of the network oscillate periodically with the same frequency and correlated phase \cite{ZAK14}; these are called \textit{amplitude chimeras}. This is in contrast, for example, to amplitude-mediated chimeras \cite{SET14}, where chimera behavior is observed with respect to both amplitude and phase. Another example of a chimera pattern where coherence (incoherence) corresponds to order (disorder) in space, but there are no oscillations in time, is \textit{chimera death}, i.e., a steady state chimera pattern found for nonlocal \cite{ZAK14} and mean-field diffusive coupling \cite{BAN15}. Recently, a chimera state which combines synchronized oscillations and steady states coexisting in space has been reported \cite{DUT15,BAN16}. Amplitude chimeras and chimera death emerge in networks with symmetry-breaking coupling, which is a crucial condition for the emergence of non-trivial steady states (oscillation death) \cite{SCH15b}. 

The amplitude chimera is defined as the
coexistence of two distinct domains separated in space: one subpopulation is oscillating with spatially coherent amplitude and the
other exhibits oscillations with spatially incoherent amplitudes, i.e., the sequence of amplitudes of neighboring oscillators is random. In the incoherent domain the oscillation amplitudes and centers of mass are uncorrelated, whereas the phases of neighboring oscillators are correlated, in contrast to classical chimeras (e.g., phase chimeras and amplitude-mediated chimeras).

It is important to note that in the amplitude chimera state individual oscillators never become chaotic: they remain periodic and thus in the incoherent domain all the oscillators are temporally periodic but spatially chaotic.  
According to a recently suggested classification amplitude chimeras are categorized as transient chimeras \cite{KEM16}. Indeed amplitude chimeras are transients towards the completely synchronized state. Their lifetime strongly depends on the initial conditions in the deterministic case. In contrast to classical chimeras  \cite{WOL11,ROS14a} or transient chaos in spatially extended systems \cite{LAI11a,TEL15}, where the transient time exponentially increases with the system size, for amplitude chimeras the transient time decreases and saturates for large system size \cite{LOO16}. Thus, the transient nature of amplitude chimeras cannot be related to a finite size effect. Amplitude chimeras can also be observed in systems with time-delay and under the impact of noise \cite{ZAK15a}. 

In the present study we investigate the phase-space structure of this chimera pattern.  Amplitude chimeras are found to represent saddles in the high-dimensional phase space of the corresponding network, i.e., they have both stable and unstable manifolds, which explains their transient nature. Our goal is to investigate amplitude chimeras in a ring of Stuart-Landau oscillators with non-local coupling and to study their stability using Floquet analysis. In particular, we calculate the real parts of the Floquet exponents for a wide range of system parameters (coupling strength and range) in the regime where amplitude chimeras exist. Positive real parts of Floquet exponents correspond to the unstable manifold of the saddle and negative ones characterize its stable manifold.  All investigations are performed for a deterministic system without noise and time-delay. Using the eigenvectors spanning the stable and unstable manifolds of the saddle cycle, we relate the amplitude chimera lifetime to the geometric structure of the phase space, and investigate the structural change with increasing coupling range, and its effect upon the lifetime.

\section{Model}
The Stuart-Landau model is the normal form of a nonlinear oscillator near a supercritical Hopf bifurcation.
We consider a ring network of $N$ Stuart-Landau oscillators \cite{KUR02a,ZAK14,ZAK15b,ZAK16}, $j\in \{1,...,N \}$, all indices modulo $N$, which are coupled with the strength $\sigma $ to their $P$ nearest neighbors in each direction ($r=P/N$ defines the dimensionsless coupling range):
\begin{equation}\label{EQ:SL_network}
\dot{z_j} =f(z_j) + \frac{\sigma}{2P} \sum_{k=j-P}^{j+P} (Re(z_k) - Re(z_j)),
\end{equation}
where 
\begin{equation}\label{EQ:SL}
f(z_j) = (\lambda + i \omega -{|z_j| }^2 ) z_j , 
\end{equation}
and $z_j = x_j+i\, y_j \!=\! r_j e ^{i\phi_j} \in \mathbb{C}$ , with $x_j,y_j,r_j,\phi_j \in \mathbb{R}$, and $\lambda, \omega > 0$. Without coupling the system undergoes a Hopf bifurcation at $\lambda \!=\! 0$, so that for $\lambda > 0$ a single Stuart-Landau oscillator performs self-sustained oscillations with frequency $\omega$ and follows the limit cycle trajectories with the radius $r_j\!=\!\sqrt{ \lambda}$, and the unique fixed point ($x_j\!=\!y_j\!=\!0$) is unstable. The periodic orbit $z(t)$ is rotationally ($S^1$) invariant.

\begin{figure}[]
\begin{center}
\includegraphics[width=0.7\linewidth]{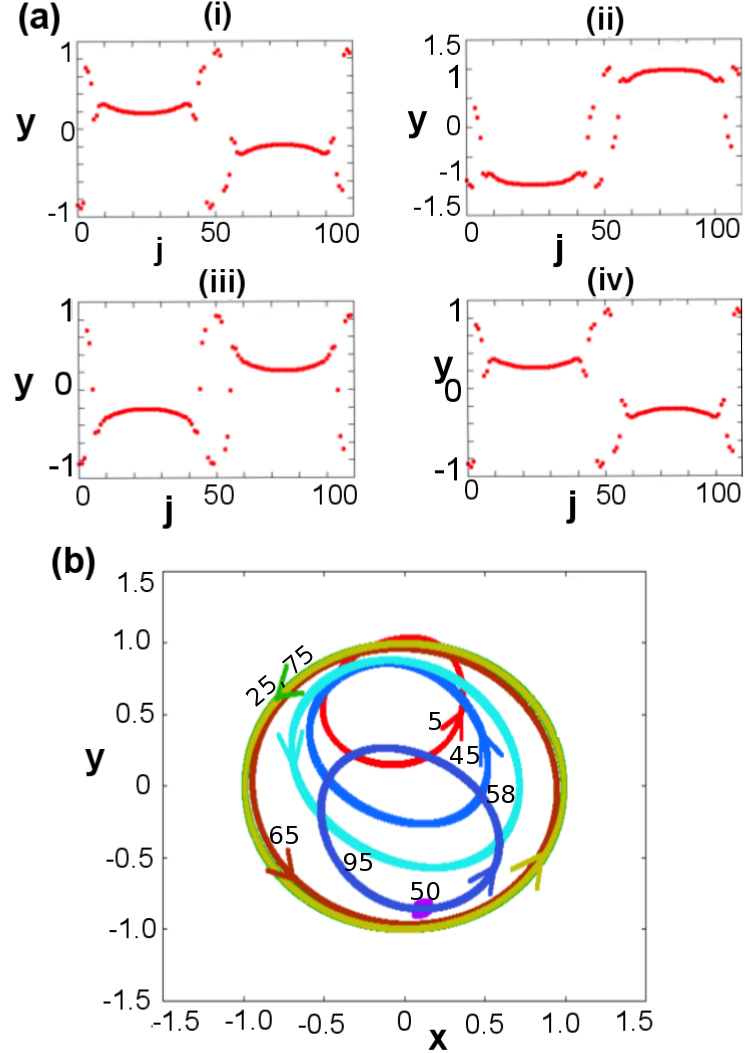}
\end{center}
\caption{Amplitude chimeras in a Stuart-Landau network of $N=100$ nodes. (a) Snapshots within one period $T=\pi$ for: (i)  $t=0$; (ii) $t=\frac{T}{4}$; (iii) $t=\frac{T}{2}$; (iv) $t=T$. (b) Phase portrait of 8 selected nodes: $j=25$, $65$, $75$ from the coherent domain and $j=5$, $45$, $50$, $58$, $95$ from the incoherent one. Other parameters: $r=0.04$, $\sigma=13$, $\lambda=1$, $\omega=2$. Initial transients of $t_0=628$ were skipped.}
\label{FIG:AC}
\end{figure}

However, the coupling term in Eq.~(\ref{EQ:SL_network}) breaks the $S^1$ symmetry, which is a crucial condition to observe nontrivial inhomogeneous steady states $z_j \ne 0$, i.e., oscillation death \cite{KOS13a,ZAK13a}.
In the following we set the bifurcation parameter $ \lambda $=1 and the oscillation frequency $ \omega $=2.
Besides steady states patterns related to oscillation death \cite{SCH15b}, the coupled system~(\ref{EQ:SL_network}) exhibits coherent in-phase synchronized oscillations or travelling waves, and partially coherent, partially incoherent amplitude chimera states, which are long-lasting transient states \cite{ZAK14,ZAK15b,LOO16}. They are presented in Fig.~\ref{FIG:AC} for a typical set of parameters.
The snapshots (a) show two coherent, synchronized domains in phase ($j=10 - 39$) and anti-phase ($j=60 - 89$), respectively,
and two incoherent domains ($j=40 - 59$, $j=90 - 9$), where the sequence of oscillations in the upper and lower $z$-plane is random. The phase portrait (b) shows that the synchronized oscillators (e.g., $j=25,65,75$) follow the limit cycle (unit circle) of the uncoupled system to a very good approximation, while the incoherent oscillators have smaller amplitudes, and their centers of mass are shifted from the origin to the upper or lower complex half-plane.

\section{Method}
The Floquet theory is a mathematical tool to study the local stability of periodic solutions, i.e., limit cycles. For a system of differential equations
\begin{equation}\label{EQ:1}
\mathbf{\dot{x}}=\mathbf{f}(\mathbf{x}(t)),
\end{equation}
with $ \mathbf{x}(t) \in {\mathbf{R}}^{n} $ assume that there exists a periodic solution $\mathbf{\chi}(t)=\mathbf{\chi}(t+T)$. Writing Eq.(\ref{EQ:SL_network}) in terms of the real variables $x_j$ and $y_j$, we have $n=2N$. We analyse the stability of the periodic orbit by considering solutions in its vicinity:
\begin{equation}\label{EQ:2}
\mathbf{x}(t)=\mathbf{\chi}(t) + \delta\mathbf{x}(t).
\end{equation}
The linearized equation
\begin{equation}\label{EQ:4}
 \delta\mathbf{\dot{x}}(t) = \mbox{D}\mathbf{f}(\mathbf{\chi}(t))\delta\mathbf{x}(t).
\end{equation}
where $ \mbox{D}\mathbf{f}(\mathbf{\chi}(t)) $ is the Jacobian, evaluated at $ \mathbf{\chi}(t) $, has the solution 
\begin{equation}\label{EQ:20}
\delta \mathbf{x}(t) = \mathbf{U}(t)\delta \mathbf{x}(0)
\end{equation}
with the initial condition $ \delta\mathbf{x}(0) $, and the fundamental matrix $ \mathbf{U}(t) $ satisfies
\begin{equation}\label{EQ:21}
\dot{\mathbf{U}}(t) = D\mathbf{f}(\mathbf{\chi}(t))\mathbf{U}(t), \hspace{1cm} \mathbf{U}(0) = \mathbf{1}.
\end{equation}
The time evolution operators obey
\begin{equation}\label{EQ:8}
\mathbf{U}(t+T) = \mathbf{U}(t)\mathbf{U}(T),
\end{equation}
and $\mathbf{U}(T)$ is called monodromy matrix. 

\begin{figure}[]
\begin{center}
\includegraphics[width=0.8\linewidth]{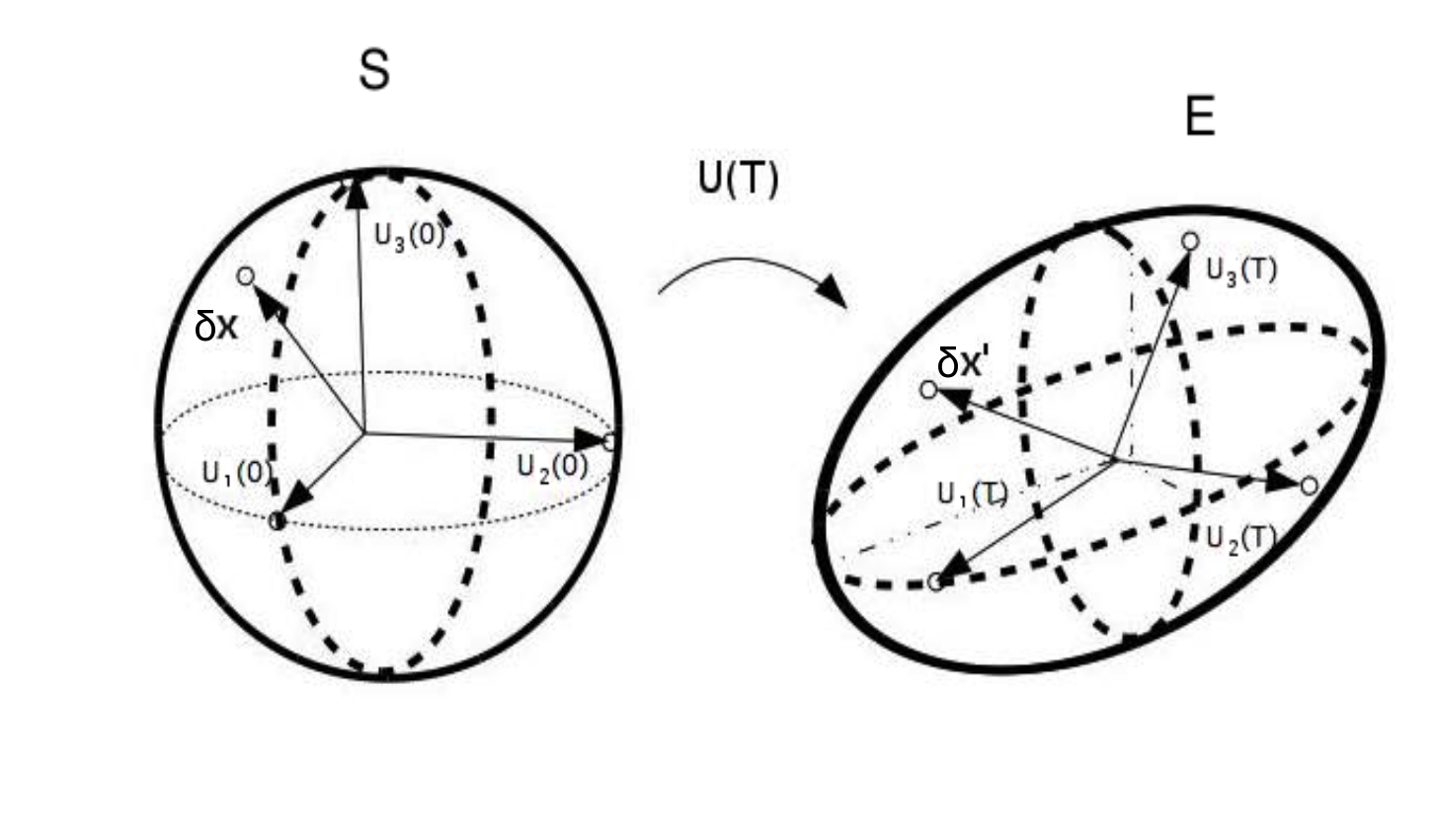}
\end{center}
\caption{Geometric interpretation of monodromy matrix $\mathbf{U}(T)$ for a three-dimensional Euclidean space (see text).}
\label{FIG:mon}
\end{figure}

In Fig.~\ref{FIG:mon} we illustrate this by a simple geometric picture (which strictly holds only for a symmetric monodromy matrix):
The initial conditions of  fundamental solutions $ \phi_{k}(0) (k=1,2,...,n)$ for $t=0$ are located on the unit sphere $S$. Each point of $S$ is mapped to the ellipsoid $E$, which contains all the fundamental solutions $ \phi_{k}(T) $ after one period $ T$ and whose principal axes correspond to the eigenvectors $\mathbf{p}_k$ and the eigenvalues $\mu_k$ (Floquet multipliers) of $\mathbf{U}(T)$:
\begin{equation}\label{EQ:ev}
\mathbf{U}(T){\mathbf{p}}_{k}={\mu}_{k}{\mathbf{p}}_{k}
\end{equation}
provides $N$ eigenvalues $\mu$, numbered by  index $k$, and corresponding eigenvectors $ \mathbf{p} $. The eigenvalues $ \mu_{k} $ of the monodromy matrix, the so-called \textit{Floquet multipliers}, characterize the stability of a period orbit $\mathbf{\chi}(t)$.
 
If all $|{\mu}_{k}| < 1$, then the ellipsoid $ E $ is located inside the sphere $S$, i.e., after one period $T$ the perturbations in all the directions decrease. If at least one $ |{\mu}_{k}| > 1 $, then the perturbations increase exponentially and the corresponding orbit is unstable. For periodic orbits $ \chi(t) $ there exists always one Floquet multiplier $|{\mu}_{k}| = 1$ (Goldstone mode), where the perturbation is along the orbit. The Floquet multipliers are related to the Floquet exponents ${\Lambda}_{k}+i{\Omega}_{k}$ by
${\mu}_{k} = \exp{({\Lambda}_{k}+i{\Omega}_{k})T}$.
The periodic orbit is stable, if all $ {\Lambda}_{k} < 0$ (except for the Goldstone mode), and unstable if at least one $ {\Lambda}_{k} > 0 $.

\begin{figure}[]
\begin{center}
\includegraphics[width=0.6\linewidth]{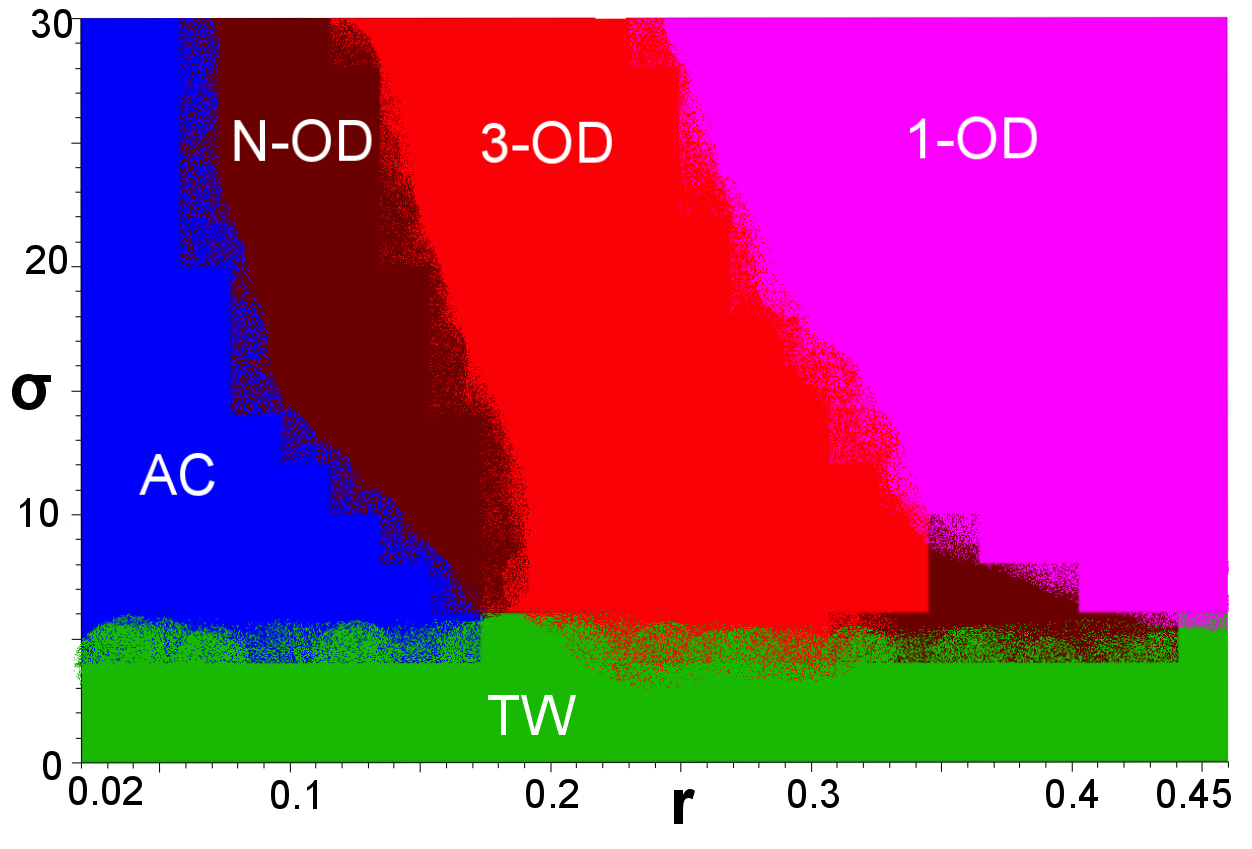}
\end{center}
\caption{Map of dynamic regimes for $N = 100$ in the plane of coupling strength $\sigma$ and coupling range $r$. 1-OD: one-cluster oscillation death; 3-OD: three-cluster oscillation death; N-OD: multi-cluster $(> 3)$ oscillation death; AC: amplitude chimera; TW: traveling wave. Other parameters: $\lambda = 1$, $\omega = 2$.} 
\label{FIG:MoR}
\end{figure}

\section{Numerical results}
\begin{figure}[]
\begin{center}
\includegraphics[width=0.40\textwidth]{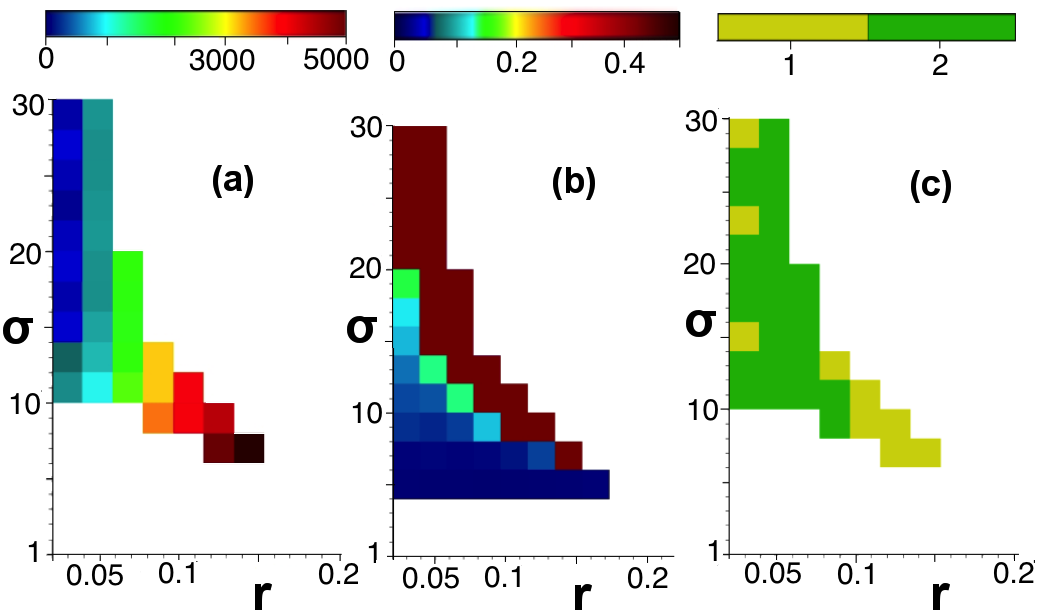}
\end{center}
\caption{(a) Transient time of amplitude chimera states in the ($\sigma, r$) plane. Maximum integration time $ t=5000 $. (b) Largest positive real parts of Floquet exponents $ \Lambda $ in the ($\sigma, r$) plane. c) Number of positive real parts of Floquet exponents $ \Lambda $ in the ($\sigma, r$) plane. Other parameters: $N=100 $, $ \lambda = 1$, $\omega = 2$.}
\label{FIG:results}
\end{figure}

Amplitude chimera states, which are long-living transients, have been found for small values of the coupling range $r$ and sufficiently strong coupling strength $\sigma$ \cite{ZAK14,ZAK15b,LOO16}. Weakly coupled systems exhibit coherent traveling waves or in-phase synchronized states. For large $\sigma$ an increase of coupling range $r$ leads to oscillation death distinguished by different numbers 
of clusters (Fig.~\ref{FIG:MoR}). 
The transient time ${t}_{tr}$ of amplitude chimeras is depicted in Fig.~\ref{FIG:results}(a) in dependence on $r$ and $\sigma$. 
In our simulations the transient time ${t}_{tr}$ is defined as the time when no oscillations with shifted center of mass are observed any more \cite{LOO16}.
Decreasing the coupling strength $\sigma$ for a fixed value of coupling range $r$ corresponds to an increase of the lifetime of the amplitude chimera. The same tendency is observed for a fixed value of $\sigma$ when increasing $r$. Dark red color denotes ${t}_{tr} >5000 $, i.e., $ {t}_{tr} > 1600T $.  Thus, amplitude chimeras are particularly long-living in networks with weak coupling and large coupling range.
Fig.~\ref{FIG:results}(b) shows the maximum real part $\Lambda$ of the Floquet exponents calculated from the monodromy matrix, which is averaged over at least 100 periods $T\approx \pi$; we do not show $\Lambda$ for states with ${t}_{tr} < 320$.
In Fig.~\ref{FIG:results}(c) we display the number of Floquet exponents with positive real parts $\Lambda$ for each set of parameters $\sigma$ and $r$. Most amplitude chimera states have one positive $\Lambda$, but some have two, corresponding to one or two unstable directions in phase space, respectively. 
For instance, increasing $\sigma$ at fixed $r=0.02$, several transitions between one and two Floquet exponents with positive real part occur; they are characterized by zero-eigenvalue bifurcations (transcritical or pitchfork bifurcations of limit cycles).
It should be noted that the second positive exponent is approximately 10 times smaller than the first one, and sometimes difficult to distinguish from the Goldstone mode which corresponds to $\Lambda=\Omega=0$, and is always present.

\begin{figure}[]
\begin{center}
\includegraphics[width=0.30\textwidth]{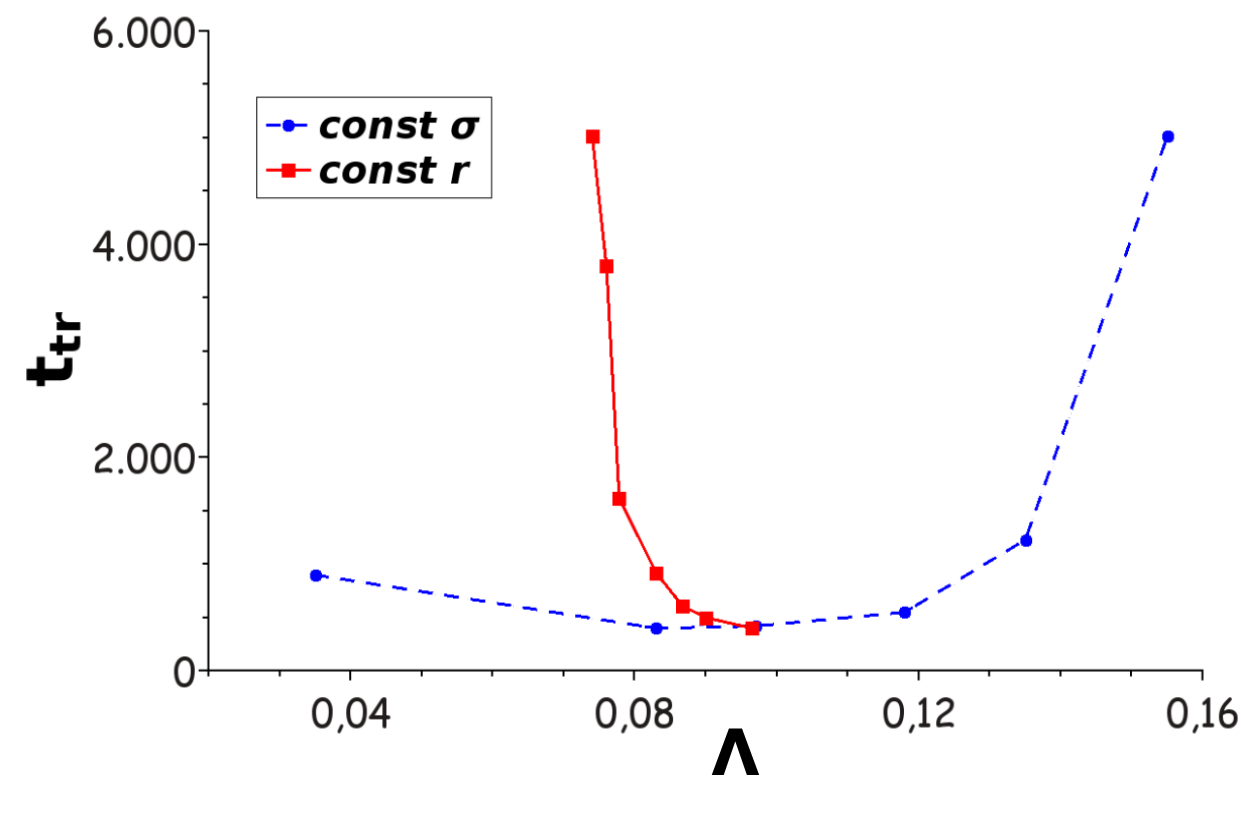}
\end{center}
\caption{Dependence of transient time $ t_{tr}$ upon real part ${\Lambda}$ of Floquet exponents (a) for constant $r=0.03$ and varying $\sigma$,  (b) for constant $\sigma=12$ and varying $r$. Parameters: $N$=100, $ \lambda$=1, $\omega$ = 2.}
\label{FIG:resb}
\end{figure}

For a fixed coupling range $r$ we observe an increase of the largest positive real part $\Lambda$ of the Floquet exponents with increasing $\sigma$ (Fig.~\ref{FIG:results}(b)). Similarly, when changing the network topology by increasing $r$ at fixed $\sigma$, we also observe that $\Lambda$ increases. Comparing Fig.~\ref{FIG:results}(a) and (b), we find that within the same network topology, i.e., fixed $r$, the transient times decrease while the positive real part of the dominant Floquet exponent increases. The escape rate $\Lambda$ from the saddle along the unstable direction increases, which leads to a shorter transient time $t_{tr} \sim \frac{1}{\Lambda}$.
However, for constant $\sigma$ with increasing $r$ both the lifetime of the amplitude chimera and $\Lambda$ increase. These two different cases are visualized by plotting $t_{tr}$ vs. $\Lambda$ in Fig.~\ref{FIG:resb} for constant $r$ (solid red curve) and constant $\sigma$ (dashed blue curve), respectively. The change in phase space structure will be elaborated below.

These results verify the hypothesis \cite{LOO16} that amplitude chimeras are saddle-orbits in phase space with a small number (one or two) of unstable dimensions.

\section{Eigenvectors}
\begin{figure}[]
\begin{center}
\includegraphics[width=0.40\textwidth]{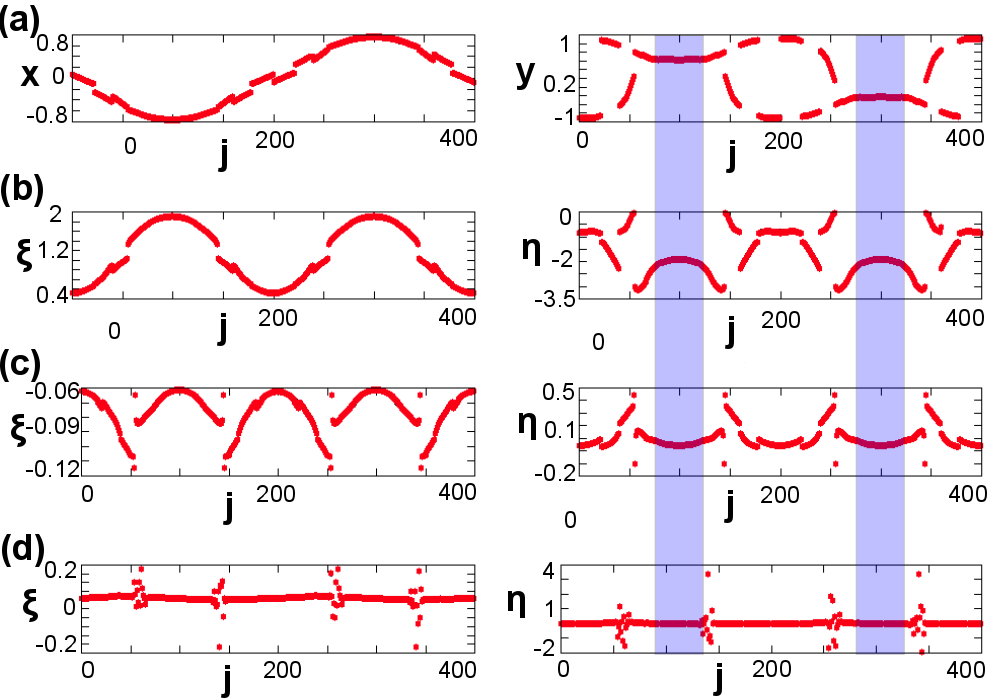}
\end{center}
\caption{Real (left column) and imaginary parts (right column) of (a) snapshot $z_j$, (b) eigenvectors associated with the largest positive $\Lambda=0.135$, (c) with the second positive $\Lambda=0.053$,
(d) with the 
first negative $\Lambda=-0.133$. 
Parameters: $N$=400, $ \lambda$ = 1, $\omega$ = 2, $\sigma$=23, $ r$=0.06.}
\label{FIG:eigenvectors}
\end{figure}

Essential information on the structure of the phase space near the saddle is given by the eigenvectors ${\mathbf{p}}_{k}$ of the monodromy matrix $\mathbf{U}(T)$ corresponding to the Floquet multipliers $\mu_k$, see Eq.~(\ref{EQ:ev}).
The eigenvectors associated with positive real parts of Floquet exponents span the unstable manifold of the saddle. 
In Fig.~\ref{FIG:eigenvectors}(b),(c) we show the two eigenvectors associated with positive real parts $\Lambda$ of the Floquet exponents,
and in (d) the eigenvector related to the leading stable Floquet exponent,  
for an amplitude chimera state in a network with $N=400$. 
For reference a snapshot of the amplitude chimera is given in Fig.~\ref{FIG:eigenvectors}(a); it illustrates the location of coherent ($j \in{(65, ..., 135)}, \in{(265, ..., 335)}$, shaded blue) and incoherent domains. The coherence and incoherence shows up most
prominently in the imaginary part ($y$-variable), i.e., the right column. Panels (b) and (c) show that the components of the eigenvector associated with the unstable manifold have much more distributed values in the incoherent domain than the nodes in the coherent domain which are almost constant. This is due to the fact that incoherent domains represent sources of instability, which push our network away from the amplitude chimera state towards the completely synchronized state. In contrast, the components of the eigenvector associated with a negative real part of the Floquet exponents (stable manifold of the amplitude chimera) are almost zero everywhere except at the boundaries between coherent and incoherent domains. 
 
\section{Phase space structure}
\begin{figure}[]
\begin{center}
\includegraphics[width=0.40\textwidth]{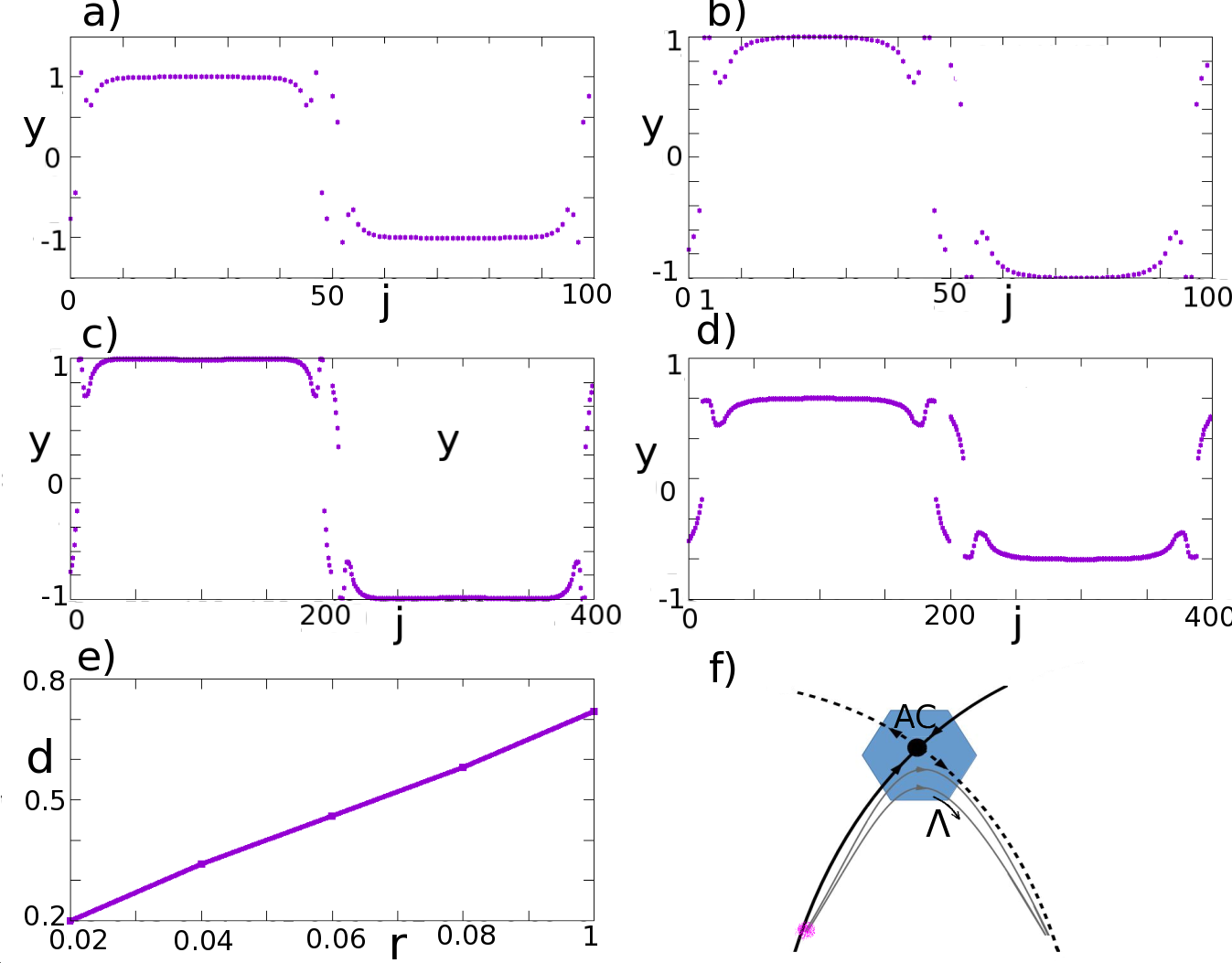}
\end{center}
\caption{Snapshots $y_j$ of amplitude chimeras for $N=100$ (a) $r=0.02$, (b) $r=0.04$, and for $N=400$ (c) $r=0.02$, (d) $r=0.04$; 
(e) width of the two incoherent domains $d$ (normalized by $N$) vs. $r$ for $N=100$; (f) Sketch of the phase space structure of amplitude chimera (AC) as a saddle orbit. 
$\Lambda$: positive real part of Floquet exponent; solid lines: stable manifold; dashed lines: unstable manifold; thin lines: possible trajectories. 
}
\label{FIG:phasespace}
\end{figure}

Based upon the eigenvectors, we will now discuss the change in phase space structure induced by varying the network topology, i.e., the coupling parameter $r$. The geometric features of phase space in the vicinity of the amplitude chimera (AC) are schematically represented in Fig.~\ref{FIG:phasespace}(f). Two trajectories approaching the amplitude chimera saddle along the stable manifold (solid lines), and then escaping along the unstable manifold (dashed lines) with escape rate $\Lambda$ are shown. 
If more neighbors are coupled to each element, i.e., for  larger $r$, this leads to an increase in the width of the incoherent domain 
(Fig.~\ref{FIG:phasespace}(a),(b)) because those synchronized elements which are located at the edge of the coherent domain experience more influence from the elements in the incoherent domain, and hence they also become desynchronized. A larger incoherent domain means that it takes longer time to reach the completely synchronized global attractor, hence the lifetime increases.  Fig.~\ref{FIG:phasespace}(c),(d) shows the corresponding snapshots for larger $N=400$. Again, the width of the incoherent domains increases with $r$, and additionally it can be seen that with increasing $N$ the width for fixed $r$ shrinks, cf. panels (a) and (c), or (b) and (d), respectively. This explains why the lifetime decreases with increasing $N$, in contrast to classical chimeras
\cite{WOL11,ROS14a}. Fig.~\ref{FIG:phasespace}(e) depicts the relative width of the two incoherent domains vs. $r$ for $N=100$ which clearly follows a linear relation.

For all parameter values we choose the same initial condition $\mathbf{x}^{AC}$ which is approximately equal to the amplitude chimera, and is constructed by simulating the system for 6500 timesteps ($t \approx 20T$) starting from a fully antisymmetric state as in \cite{LOO16}: the first half of the nodes $j \in{(1, ..., N/2)}$ is set to $({x}_{j},{y}_{j}) = (-1, 1)$, whereas the second half $j \in{(N/2 +1, ..., N)}$ is set to$({x}_{j},{y}_{j}) = (1, -1)$. The observation of amplitude chimera states is related with the symmetry
of the initial condition. While the chimera itself is a saddle in phase space, the dominant unstable direction (given by the symmetric eigenvector corresponding to the eigenvalue with largest positive real part) is orthogonal to antisymmetric initial conditions. Normally long transients occur which are caused by tiny violations of the symmetry of the system during numerical integration. However, on changing the parameters as demonstrated above, a second small unstable eigenvalue may occur and the corresponding unstable direction lies within the subspace of antisymmetric initial condions. This small subleading eigenvalue $\Lambda_2$ then causes transient behavior which is in agreement with the observed dependence of the transient time $t_{tr} \sim 1/\Lambda_2$ on the system parameters, since 
$\Lambda_2$ decreases with increasing width of the incoherent domain and hence with $r$. 
\\

\section{Amplitude chimeras in small networks}
\begin{figure}[]
\begin{center}
\includegraphics[width=0.40\textwidth]{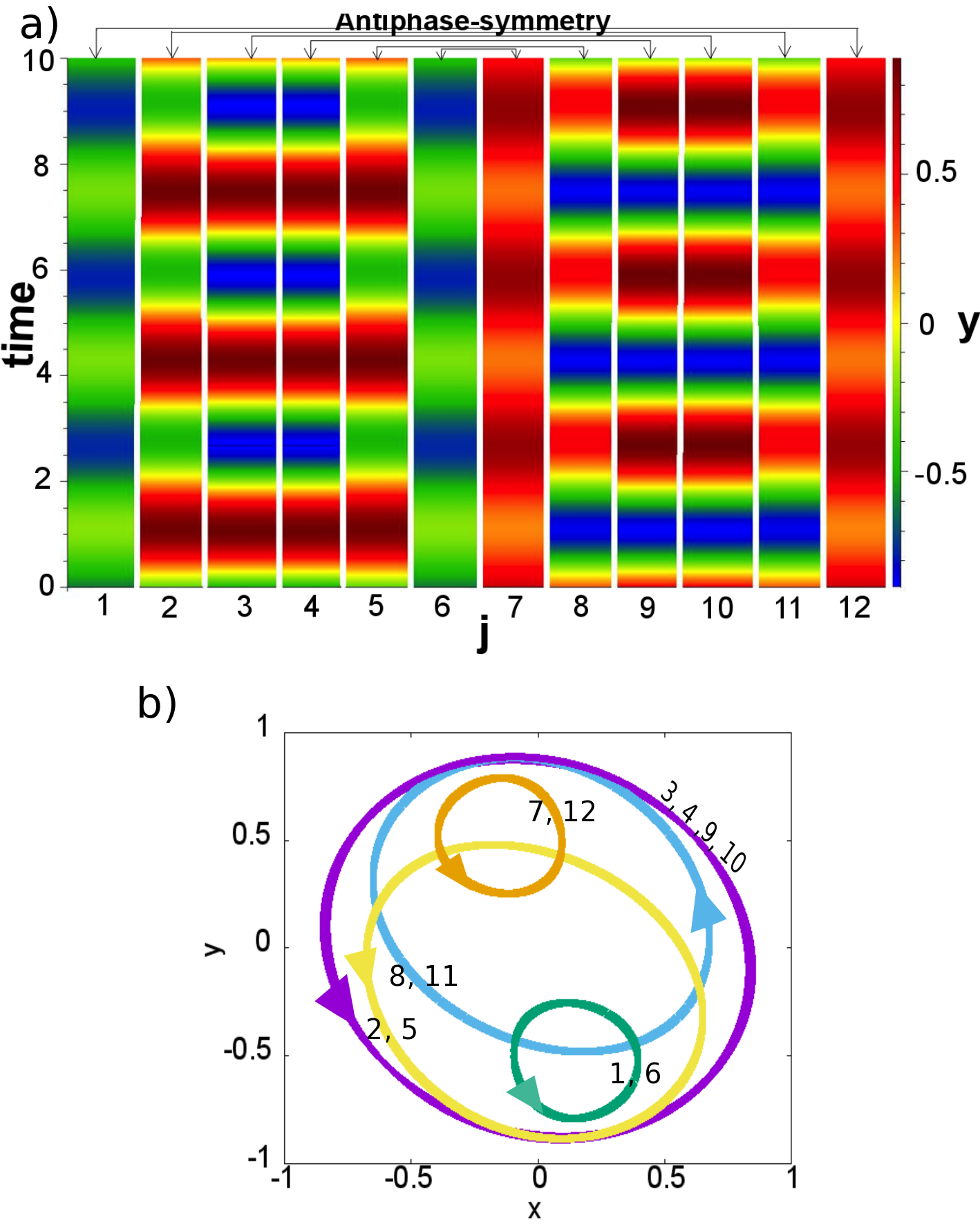}
\end{center}
\caption{(a) Space-time plot of an amplitude chimera state ${y}_{j}(t)$ in a very small network. (b) Phase portrait. Parameters: $N = 12$, $\lambda = 1$, $\omega = 2$, $r = \frac{1}{12}$, $\sigma = 5$.}
\label{FIG:12}
\end{figure}

Small networks are more amenable to analytical and numerical investigations, however, it becomes more difficult to clearly distinguish domains of coherence and incoherence. A minimum example is shown in Fig.~\ref{FIG:12} for $N=12$ nodes. Here amplitude chimeras are found only for  $P=1$ ($r=1/12$). Two coherent domains with elements $j =3,4$ and $j=9,10$, respectively, can be seen. Between these domains there are incoherent domains where the oscillation amplitude decreases, and the center of mass is shifted from the origin, see the phase portraits in Fig.~\ref{FIG:12}(b). Note that the anti-phase symmetry between elements $j=1,...,6$ and $j= 7,...,12$ is observed. The calculation of the three dominant Floquet multipliers for this set of coupling parameters gives $\Lambda=0.363$, $\Lambda= -0.359$, and the Goldstone mode $\Lambda \approx 0.059$, in agreement with our previous analysis. 

\section{Conclusions}
We have studied the stability of amplitude chimera states in Stuart-Landau oscillator networks using Floquet theory, and established the phase-space structure of these long-living transient states as saddles with one or two repelling directions given by the Floquet exponents with positive real parts and the corresponding eigenvectors. Amplitude chimeras are particularly long-living (thousands of periods) in weakly coupled networks or in networks whose topology is characterized by a large number of nearest neighbours. We have explained the behavior of the transient times in dependence on the strength $\sigma$ and the range $r$ of the nonlocal coupling by the changes in phase space structure. In particular, the increase of the chimera lifetime with increasing coupling range at fixed coupling strength results from the increase in the width of the incoherent domain. For antisymmetric initial conditions, the escape rate from the saddle is determined by numerical deviations from strict antisymmetry or by the second unstable eigenvector, whose eigenvalue has a real part $\Lambda_2$ which decreases with increasing $r$. 
At the same time, the positive real part $\Lambda_1$ of the leading Floquet exponent increases with $r$, but has no influence upon the chimera lifetime since its direction is orthogonal in phase space. \\

If the coupling strength $\sigma$ is increased at fixed coupling range $r$, the positive real part $\Lambda$ of the Floquet exponents increases, and the chimera lifetime ${t}_{tr}$ decreases inversely proportionally to $\Lambda$. In summary, amplitude chimera states with small coupling strength $\sigma$ and large coupling range $r$ have the longest lifetimes.\\ 

We have also presented simulations of amplitude chimera states in a minimum network of $N=12$ elements, and showed that these are saddles with one unstable direction. \\

\acknowledgments
This work was supported by DFG in the framework of SFB 910.

\bibliography{ref}  
\bibliographystyle{prsty-fullauthor}

\end{document}